\

$${\bf Bell \ inequality, \ nonlocality \ and \ analyticity}$$

$$M. \ Socolovsky $$

\centerline {\it Departamento de F\'\i sica Te\'orica, Universidad de Valencia, Burjassot 46100, Espa$\tilde{n}$a}

\centerline{\it and}

\centerline {\it Instituto de Ciencias Nucleares, Universidad Nacional Aut\'onoma de M\'exico}

\centerline{\it Circuito Exterior, Cd. Universitaria, 04510, M\'exico D.F., M\'exico}

\

The Bell and the Clauser-Horne-Shimony-Holt inequalities are shown to hold for both the cases of complex and real analytic nonlocality in the setting parameters of Einstein-Podolsky-Rosen-Bohm experiments for spin ${{1}\over{2}}$ particles and photons, in both the deterministic and stochastic cases. Therefore, the theoretical and experimental violation of the inequalities by quantum mechanics excludes all hidden variables theories with that kind of nonlocality. In particular, real analyticity leads to negative definite correlations, in contradiction with quantum mechanics.

\


\


\

{\bf 1. Introduction}

\

In a recent paper, in the context of hidden variables theories, Fahmi [1] derived the deterministic Clauser-Horne-Shimony-Holt (CHSH) inequality [2], from which the Bell inequality is easily obtained [3], allowing certain conditions of nonlocality in the parameters specifying the orientation of the Stern-Gerlach ({\it SG}) apparatus or polarizers in an EPRB experiment [4,5]. Since the inequalities are violated by quantum mechanics both theoretically [2,3] and experimentally [6,7], Fahmi concluded that a broader class of deterministic hidden variables theories are excluded, namely those obeying Bell locality [3] and those which being non local, are of the type that he considered. 

It arises the question as to which is the most general form of nonlocality that allows to obtain the Bell and/or the CHSH inequalities. In this paper we investigate the particular case of nonlocalities which are expressable as real valued {\it analytic functions} of the setting parameters, both in the cases that these parameters are considered complex or real valued quantities. In the complex case the proof is simple and it is based on the fact that as a consequence of the Cauchy-Riemann (CR) equations, a real valued complex analytic function with connected domain is a constant. The argument for using complex parameters is based on the fact that the set of unit vectors {\bf a} and {\bf b} specifying the directions of two SG apparatus consists of the product of two spheres,  $S^2\times S^2$, each of which, through the stereographic  projection is equivalent to the Riemann sphere or complex projective line: $P ^1_c=C \cup \{\infty\}$; as is well known, for spin ${{1}\over{2}}$ non relativistic quantum particles, $P^1_c$ is a natural space to describe their physics [8]. Also, several authors [9-11] have pointed out the {\it necessary} complex character of quantum mechanics, which we assume here for any of its extensions. So, one is naturally guided to the use of complex valued setting parameters.

The case of real analytic nonlocality is nevertheless analized, and it is also shown to be violated by quantum mechanics, this time at the level of the correlations functions, since it produces for them a negative definite result.

The analysis is extended to the non deterministic stochastic case, and therefore our conclusion is that {\it if we assume the existence of hidden variables, then all deterministic hidden variables theories as well as all stochastic hidden variables theories with complex or real analytic nonlocality in the detecting settings (SG) are excluded.}
 
\

{\bf 2. Complex analytic nonlocality in Bell theorem}

\

Let 1 and 2 be the Einstein-Podolsky-Rosen-Bohm-Bell [3-5] pair of particles, that is, 1 and 2 is a system of two quantum particles which at the moment of their preparation are in the singlet state for the spin ${{1}\over{2}}$ case, or in a state of opposite polarization for the photon case; and at the moment of their detection they are separated by a space-like interval. The state of the particles is assumed to be described by the wave function $\psi$ and by a set of unspecified real quantities $\lambda=(\lambda_1,...,\lambda_r)\in \Lambda$, the so called hidden variables, classically distributed with a non negative normalized probability density $\rho(\lambda)$ such that $\int_{\Lambda}d\lambda \rho(\lambda)=1$. (We recall that a {\it hidden variables theory} is an {\it extension} of ordinary quantum mechanics, namely, it does not replace $\psi$ by $\lambda$ but adds $\lambda$ to $\psi$.) The quantities which are measured, $A$ and $B$, for the particles 1 and 2 respectively, and which take values in the set $\{+1,-1\}$ ({\it determinism}), are functions of the unit vectors {\bf a} and {\bf b} in 3-space $R^3$ (the directions of the SG apparatus for spin ${{1}\over{2}}$ or polarizers in the case of photons), of the wave function $\psi$, and of the hidden variables $\lambda$: $$A=A({\bf a},{\bf b};\psi, \lambda), \ B=B({\bf a},{\bf b}; \psi, \lambda).$$ In the original derivation by Bell [3], his strong locality hypotesis [12-14] says that $A$ depends on {\bf a} but not on {\bf b}, and viceversa: $B$ depends on {\bf b} but not on {\bf a}. The average over $\lambda$ of the product of $A$ and $B$ measures the {\it statistical} correlation between the particles, and is given by $$P({\bf a}, {\bf b}; \psi)=\int_{\Lambda}d\lambda \rho(\lambda)A({\bf a},{\bf b}; \psi, \lambda)B({\bf a},{\bf b};\psi,\lambda). \eqno{(1)}$$ 
(The statistical character of the correlation is expressed in the fact that  $\vert P({\bf a},{\bf b};\psi) \vert \leq 1$, while a perfect correlation corresponds to $\vert P({\bf a},{\bf b}; \psi) \vert =1$, as in EPRB with {\bf b}={-\bf a}.)
{\bf a} and {\bf b}, being unit vectors in 3-space, are the elements of the 2-sphere, and so $({\bf a},{\bf b})\in S^2\times S^2$. Through the stereographic projection, $S^2\times S^2$ is homeomorphic to $M\equiv P^1_c\times P^1_c=(C\cup \{\infty\})\times (C\cup \{\infty\})$, which is a 2-dimensional compact connected complex manifold. In a natural way, $A$ and $B$ can be considered complex analytic functions on $M$ with {\bf a} and {\bf b} replaced by  complex numbers $z_1$ and $z_2$, and the point at infinity $\infty$. Using the Cauchy-Riemann equations [15] one can show that the fact that $A$ and $B$ take real values implies that they are constant functions of {\bf a} and {\bf b} (or $z_1$ and $z_2$), for fixed values of $\psi$ and $\lambda$, and, by continuity, have the same values on the 1-dimensional complex submanifold $\{({\bf a},\infty), (\infty,{\bf b}), (\infty,\infty)\}_{({\bf a},{\bf b})\in S^2\times S^2}$. Then the power series for $A$ and $B$ are respectively given by $$A({\bf a},{\bf b};\psi,\lambda)=\alpha_{{\bf a}{\bf b}}(\psi,\lambda), \ B({\bf a},{\bf b};\psi,\lambda)=\beta_{{\bf a}{\bf b}}(\psi,\lambda)$$ where for fixed $\lambda$ the r.h.s.'s are independent of $({\bf a},{\bf b})\in S^2\times S^2$. 

Proceeding as usual for the difference of correlation functions, we have:
$$P({\bf a},{\bf b};\psi)-P({\bf a},{\bf b}^\prime;\psi)=\int_\Lambda d\lambda \rho (\lambda)\alpha_{{\bf a}{\bf b}}(\psi,\lambda)\beta_{{\bf a}{\bf b}}(\psi,\lambda)-\int_\Lambda d\lambda \rho(\lambda)\alpha_{{\bf a}{\bf b}^\prime}(\psi,\lambda)\beta_{{\bf a}{\bf b}^\prime}(\psi,\lambda)$$ $$=\int_\Lambda d\lambda \rho (\lambda)(\alpha_{{\bf a}{\bf b}}(\psi,\lambda)\beta_{{\bf a}{\bf b}}(\psi,\lambda)(1\pm \alpha_{{\bf a}^\prime {\bf b}^\prime}(\psi,\lambda)\beta_{{\bf a}^\prime {\bf b}^\prime}(\psi,\lambda))$$ $$-\alpha_{{\bf a}{\bf b}^\prime}(\psi,\lambda)\beta_{{\bf a}{\bf b}^\prime}(\psi,\lambda)(1\pm \alpha_{{\bf a}^\prime {\bf b}}(\psi,\lambda)\beta_{{\bf a}^\prime {\bf b}}(\psi,\lambda)))$$ $$=P({\bf a},{\bf b};\psi)-P({\bf a},{\bf b}^\prime;\psi)\pm \int_\Lambda \rho(\lambda)I(\psi,\lambda)$$ with $$I(\psi,\lambda)=\alpha_{{\bf a}{\bf b}}(\psi,\lambda)\beta_{{\bf a}{\bf b}}(\psi,\lambda)\alpha_{{\bf a}^\prime {\bf b}^\prime}(\psi,\lambda)\beta_{{\bf a}^\prime {\bf b}^\prime}(\psi,\lambda)-\alpha_{{\bf a}{\bf b}^\prime}(\psi,\lambda)\beta_{{\bf a}{\bf b}^\prime}(\psi,\lambda)\alpha_{{\bf a}^\prime {\bf b}}(\psi,\lambda)\beta_{{\bf a}^\prime {\bf b}}(\psi,\lambda).$$ 

Since $$\alpha_{{\bf a}{\bf b}}(\psi,\lambda)=\alpha_{{\bf a}^\prime {\bf b}^\prime}(\psi,\lambda)=\alpha_{{\bf a}^\prime {\bf b}}(\psi,\lambda)=\alpha_{{\bf a}{\bf b}^\prime}(\psi,\lambda)=const. \in \{+1,-1\}$$ for each $(\psi,\lambda)$, and analogously for the $\beta's$, then $$I=0,$$ and we can proceed straightforwardly to obtain the CHSH inequality (2): $$\vert P({\bf a},{\bf b};\psi)-P({\bf a},{\bf b}^\prime;\psi)\vert \leq \int_\Lambda d\lambda \rho(\lambda)(\vert 1\pm \alpha_{{\bf a}^\prime {\bf b}^\prime}(\psi,\lambda)\beta_{{\bf a}^\prime {\bf b}^\prime}(\psi,\lambda)\vert +\vert 1 \pm \alpha_{{\bf a}^\prime {\bf b}}(\psi,\lambda)\beta_{{\bf a}^\prime {\bf b}}(\psi,\lambda)\vert )$$ $$=2 \pm (\int_\Lambda d\lambda \rho (\lambda)\alpha_{{\bf a}^\prime {\bf b}^\prime}(\psi,\lambda)\beta_{{\bf a}^\prime {\bf b}^\prime}(\psi,\lambda)+\int_\Lambda d\lambda \rho (\lambda)\alpha_{{\bf a}^\prime {\bf b}}(\psi,\lambda)\beta_{{\bf a}^\prime {\bf b}}(\psi,\lambda))$$ $$=2\pm (P({\bf a}^\prime, {\bf b}^\prime;\psi)+P({\bf a}^\prime,{\bf b}; \psi))\leq 2 \pm \vert P({\bf a}^\prime, {\bf b}^\prime;\psi)+P({\bf a}^\prime, {\bf b};\psi)\vert$$ which implies $$\vert P({\bf a},{\bf b};\psi)-P({\bf a},{\bf b}^\prime; \psi)\vert + \vert P({\bf a}^\prime,{\bf b}^\prime;\psi)+P({\bf a}^\prime, {\bf b};\psi)\vert \leq 2. \eqno{(2)}$$

\

{\bf 3. The stochastic case}

\

The stochastic case, in which the values of $A$ and $B$ in $\{+1,-1\}$ are replaced by average values $\bar{A}$ and $\bar{B}$ in the interval [-1,+1], is easily obtained from the previous result, by first deriving a formal mathematical inequality and then by making use of the Bell factorization hypotesis [16], with nonlocality incorporated as before as complex analytic dependence on the parameters of the measuring settings. 

For any $x\in R$ (or $C$), if $\vert x \vert=1$ then $\vert x \vert \leq 1$; in particular, then, $\vert A({\bf a},{\bf b};\psi,\lambda)\vert$, $\vert B({\bf a},{\bf b};\psi,\lambda)\vert   \leq 1.$ In the present case, $A, B \in R$, and the analysis carrying from eq. (1) to eq. (2) proceeds in a straightforward manner under the same assumption of complex analytic dependence of $A$ and $B$ on {\bf a} and {\bf b}, for given $\psi$ and $\lambda$.

Let $P(\mu,\nu; {\bf a},{\bf b};\psi, \lambda)\in [0,1]$ be the {\it probability} that the measurements of the spins (or polarizations) on particles 1 and 2, respectively along the directions {\bf a} and {\bf b}, give the results $\mu$ for 1 and $\nu$ for 2, for given $\psi$ and $\lambda$, with $\mu, \nu \in \{+1,-1\}$. The crucial hypotesis is that of {\it factorization} [16] of $P$: $$P(\mu,\nu;{\bf a},{\bf b};\psi,\lambda)=P_1(\mu; {\bf a},{\bf b};\psi,\lambda)P_2(\nu;{\bf a},{\bf b};\psi,\lambda). \eqno{(3)}$$ In the Bell's local version, $P_1$ depends on {\bf a} but not on {\bf b}, and viceversa for $P_2$. Clearly, $0\leq P_1,P_2 \leq 1$. The average of the above probability over the fluctuations of $\lambda$ is given by $$P(\mu,\nu;{\bf a},{\bf b};\psi)=\int_\Lambda d\lambda \rho(\lambda)P(\mu,\nu;{\bf a},{\bf b};\psi,\lambda)=\int_\Lambda d\lambda \rho(\lambda)P_1(\mu;{\bf a},{\bf b};\psi,\lambda)P_2(\nu;{\bf a},{\bf b};\psi,\lambda).$$ As in section 2, the correlation function is defined as the average of the product of the two spins (or polarizations): $$P({\bf a},{\bf b};\psi)=P(+,+;{\bf a},{\bf b};\psi)+P(-,-;{\bf a},{\bf b};\psi)-P(+,-;{\bf a},{\bf b};\psi)-P(-,+;{\bf a},{\bf b};\psi)$$ with $$P(+,+;{\bf a},{\bf b};\psi)=\int_\Lambda d\lambda \rho(\lambda)P_1(+;{\bf a},{\bf b};\psi,\lambda)P_2(+;{\bf a},{\bf b};\psi,\lambda),$$ $$P(-,-;{\bf a},{\bf b};\psi)=\int_\Lambda d\lambda \rho(\lambda)P_1(-;{\bf a},{\bf b};\psi,\lambda)P_2(-;{\bf a},{\bf b};\psi,\lambda),$$ $$P(+,-;{\bf a},{\bf b};\psi)=\int_\Lambda d\lambda \rho(\lambda)P_1(+;{\bf a},{\bf b},\psi,\lambda)P_2(-;{\bf a},{\bf b},\psi,\lambda),$$ $$P(-,+;{\bf a},{\bf b};\psi)=\int_\Lambda d\lambda \rho(\lambda)P_1(-;{\bf a},{\bf b};\psi,\lambda)P_2(+;{\bf a},{\bf b};\psi,\lambda);$$ then $$P({\bf a},{\bf b};\psi)=\int_\Lambda d\lambda \rho(\lambda)(P_1(+;{\bf a},{\bf b};\psi,\lambda)P_2(+;{\bf a},{\bf b};\psi,\lambda)+P_1(-;{\bf a},{\bf b};\psi,\lambda)P_2(-;{\bf a},{\bf b};\psi,\lambda)$$ $$-P_1(+;{\bf a},{\bf b};\psi,\lambda)P_2(-;{\bf a},{\bf b};\psi,\lambda)-P_1(-;{\bf a},{\bf b};\psi,\lambda)P_2(+;{\bf a},{\bf b};\psi,\lambda))$$ $$=\int_\Lambda d\lambda \rho(\lambda)(P_1(+;{\bf a},{\bf b};\psi,\lambda)-P_1(-;{\bf a},{\bf b};\psi,\lambda))(P_2(+;{\bf a},{\bf b};\psi,\lambda)-P_2(-;{\bf a},{\bf b};\psi,\lambda))$$ $$:=\int_\Lambda d\lambda \rho(\lambda)<A({\bf a},{\bf b};\psi,\lambda)><B({\bf a},{\bf b};\psi,\lambda)> \eqno{(4)}$$ where $<A({\bf a},{\bf b};\psi,\lambda)>$ and $<B({\bf a},{\bf b};\psi,\lambda)>$, both in the interval [-1,+1], are the {\it average values} of the spins (or polarizations) of particles 1 and 2, respectively, along the directions {\bf a} and {\bf b}. Assuming the complex analytic dependence of these quantities on {\bf a} and {\bf b}, and using the formal mathematical result at the beginning of this section, we obtain the CHSH inequality for the stochastic case, which is again given by eq. (2) but with the $P's$ given now by eq. (4).

\

{\bf 4. Real analyticity}

\

It can be easily shown that if we assume real analytic expansions for the functions $A$ and $B$ of section 2, then the correlation functions (1) become negative definite, in clear contradiction with the quantum formula (7) of section 5. In fact, let $A$ and $B$ be given by the convergent series $$A({\bf a},{\bf b}; \psi,\lambda)=\sum^{\infty}_{i,j=1}\sum^3_{r,s=1}\alpha^{rs}_{ij}a^i _rb^j_s$$ and $$B({\bf a},{\bf b};\psi, \lambda)=\sum^{\infty}_{k,l=1}\sum^3_{t,u=1}\beta^{tu}_{kl}a^k_tb^l_u$$ with $\alpha^{rs}_{ij}$ and $\beta^{tu}_{kl}$ real coefficients (functions of $\psi$ and $\lambda$). (In the real case the CR equations do not hold and the power series expansions do not reduce to constant terms.) For {\bf b}={\bf a}, (1) gives the perfect correlation $$-1=P({\bf a},{\bf a};\psi)=\int_{\Lambda}d\lambda\rho(\lambda)A({\bf a},{\bf a};\psi,\lambda)B({\bf a},{\bf a};\psi,\lambda)$$ which implies $$\int_{\Lambda}d\lambda\rho(\lambda)(A({\bf a},{\bf a};\psi,\lambda)B({\bf a},{\bf a};\psi,\lambda)+1)=0$$ and then $$B({\bf a},{\bf a};\psi,\lambda)=-A({\bf a},{\bf a};\psi,\lambda)$$ for all $\lambda$, except for a possible set of measure zero in $\Lambda$. So, $$A({\bf a},{\bf a};\psi,\lambda)=\sum^{\infty}_{i,j=1}\sum^3_{r,s=1}\alpha^{rs}_{ij}a^i_ra^j_s=\sum^{\infty}_{i,j=1}\sum^3_{r,s=1}(-\beta^{rs}_{ij})a^i_ra^j_s$$ {\it i.e.} $$\sum^{\infty}_{i,j}\sum^3_{r,s=1}(\alpha^{rs}_{ij}+\beta^{rs}_{ij})a^i_ra^j_s=0$$ which holds for all unit vectors ${\bf a}\in R^3$. Then $$\beta_{ij}^{rs}=-\alpha_{ij}^{rs} \eqno{(5)}$$ and therefore $$P({\bf a},{\bf b};\psi)=-\int_{\Lambda}d\lambda\rho(\lambda)(A({\bf a},{\bf b};\psi,\lambda))^2=-1, \eqno{(6)}$$ which is a negative constant value for the correlation function.

\

{\bf 5. Quantum correlations}

\

For completeness, and to emphasize the difference between standard quantum mechanics and its extension with hidden variables, we consider the {\it quantum correlation} between two spin ${{1}\over {2}}$ particles in the singlet state, given by the well known formula [10,17] $$P_Q({\bf a},{\bf b};\psi)=-{\bf a}\cdot {\bf b}. \eqno{(7)}$$ $P_Q$ is a real valued real analytic function of {\bf a} and {\bf b} considered as real vectors in $R^3$; however, since we are now in the framework of pure quantum mechanics, this {\it does not} contradict the main conclusion in the Introduction since hidden variables are absent. 

On the other hand, considered as a real valued {\it complex} function on the connected space $P^1_c\times P^1_c$, it can be shown to be {\it non analytic} (and therefore not necessarily a constant): in fact, using the stereographic projection $$\Phi:C\cup \{\infty\}\to S^2, \ \Phi(z)={{((z+\bar{z}),i(\bar{z}-z),1-\vert z \vert^2)}\over {1+\vert z \vert ^2}}, \ \Phi(\infty)=(0,0,-1),$$ and identifying {\bf a} with $z$ or $\infty$ and {\bf b} with $w$ or $\infty$, we obtain $$P_Q(z,\bar{z};w,\bar{w};\psi)=-{{(z+\bar{z})(w+\bar{w})-(\bar{z}-z)(\bar{w}-w)+(1-\vert z \vert ^2)(1-\vert w \vert ^2)}\over {(1+\vert z \vert ^2)(1+\vert w \vert ^2)}},$$ $$P_Q(z,\bar{z};\infty;\psi)=P_Q(\infty;z,\bar{z})={{1-\vert z \vert ^2}\over {1+\vert z \vert ^2}}, \ \ and \ \ P_Q(\infty,\infty;\psi)=-1.$$ The dependence of $P_Q$ on both $z$ and $\bar{z}$ and on $w$ and $\bar{w}$ shows that $P_Q$ is non analytic. In this form, $P_Q$ has the same non analytic character of the correlation function in the case of quantum mechanics extended with hidden variables; clearly, the agreement is not necessary since we are dealing with different theories, however, as mentioned above, it supports the view of quantum mechanics as a necessarily complex theory [9,11].

\

{\bf Acknowledgments}

\

The author thanks the Spanish Ministry of Education and Culture for a sabbatical grant. Also, the author thanks Professors Carmen Romero and Carlos A. Garc\'\i a Canal for useful discussions. This work has been partially supported by the research grant BFM2002-03681 from the Ministerio de Ciencia y Tecnolog\'\i a and from EU FEDER funds.

\

{\bf References}

\

\noindent[1] A. Fahmi, Phys. Lett. A 303 (2002)1. 

\noindent[2] J. S. Bell, Physics 1 (1964)195.

\noindent[3] J. F. Clauser, M. A. Horne, A. Shimony, R. A. Holt, Phys. Rev. Lett. 23 (1969)880.

\noindent[4] A. Einstein, B. Podolsky, N. Rosen, Phys. Rev. 47 (1935)777.

\noindent[5] D. Bohm, Quantum Theory, Prentice-Hall, 1951.

\noindent[6] A. Aspect, P. Grangier, G. Roger, Phys. Rev. Lett. 49 (1982)91.

\noindent[7] G. Weihs, T. Jennewein, C. Simon, H. Weinfurter, A. Zeilinger, Phys. Rev. Lett. 81 (1998)5039.

\noindent[8] M. Socolovsky, Advances in Applied Clifford Algebras 11(1) (2001)109.

\noindent[9] P. A. M. Dirac, The principles of Quantum Mechanics, p.18, Oxford Univ. Press, London, 1958.

\noindent[10] J. J. Sakurai, Modern Quantum Mechanics, p.10, Addison-Wesley, Reading, Mass., 1994.

\noindent[11] R. Penrose, Shadows of the Mind, p.237, Oxford Univ. Press, New York, 1994.

\noindent[12] J. P. Jarrett, No$\hat{u}$s 18 (1984)569.

\noindent[13] L. E. Ballentine, J. P. Jarrett, Am. Jour. Phys. 55(8) (1987)696.

\noindent[14] A. Shimony, Events and Processes in the Quantum World, in Quantum Concepts in Space and Time, ed. by R. Penrose and C. J. Isham, Clarendon Press, Oxford, 1986.

\noindent[15] V. S. Vladimirov, Methods of the Theory of Functions on Many Complex Variables, The M.I.T. Press, 1966.

\noindent[16] J. S. Bell, Introduction to the Hidden-Variable Question, in Proceedings of the International School of Physics, Enrico Fermi, Course IL, Varena, 1970, New York, Academic, 1971.

\noindent[17] M. Socolovsky, Rev. Mex. F\'\i s. 48 (2002)384.

\end

In a recent paper, Aguilar and Socolovsky $^{1}$ have studied geometrical and topological aspects of the abelian Aharonov-Bohm (A-B) $^{2}$ effect.  This is a gauge invariant non local quantum effect that geometrically can be thought to be induced by a non trivial flat connection on a product bundle over a space with a non trivial topology. In particular it was determined that the principal bundle relevant for the A-B effect is the product bundle $\xi _{A-B}:U(1)\to
R^{2*}\times U(1)\buildrel \ {\pi} \over \longrightarrow R^{2*}$ (for scalar particles, the wave function is a section of the associated vector bundle $\xi_C:C \longrightarrow R^{2*}\times C \buildrel \ {\pi} \over \longrightarrow R^{2*}$), where $R^{2*}\times U(1)$, the total space of $\xi_{A-B}$, is homeomorphic to an open solid 2-torus minus a circle.  The moduli space of flat connections, that is, the set of gauge equivalence classes of flat connections ${\cal M}_0={\cal C}_0/{\cal G}$, where ${\cal C}_0$ is the set of flat connections and ${\cal G}$ is the gauge group of the bundle, was shown to be isomorphic to the circle $S^1$; finally, the holonomy groups of these connections in terms of $\rho \in [0,1)$ were shown to be either the cyclic groups $H(\rho)=Z_q$, for $\rho =p/q \in Q$, or the integers $Z$, the infinite cyclic group, for $\rho \not\in Q$.

\

These geometrical properties of the A-B effect are independent of the ideal case considered here, namely that of an infinitesimally thin solenoid carrying the magnetic flux $\Phi$. In a real situation, the base space of the bundle is the plane minus a disk, which is topologically equivalent to $R^{2*}$. 

\

The result about ${\cal M}_0$ in Ref. 1, was obtained as a corollary of a general result which is valid for product bundles over any manifold $M$.  This result shows that the A-B effect is caused by the non trivial topology of M.  Here we find ${\cal M}_0$ {\it i.e.}, the moduli space for the case $M=R^{2*}$ in a simpler way. The result coincides with the previous derivation, but the explicit inclusion of the coupling constant {\it i.e.} the electric charge $e$, leads to the
result that the length of the circle is $\vert e \vert$. In section 2 we describe the gauge group ${\cal G}$, and in section 3 we rederive ${\cal M}_0$. Section 4 is a remark on the relation of this length with other fundamental lengths.

\

\

\

\noindent
---------------------------------

Corresponding author: M. S., Universidad de Valencia, telephone number: (34) 963 54 4349, fax number: (34) 963 98 3381, e-mail: miguel.socolovsky@uv.es

\

\

\

{\bf 2. The gauge group}

\

Since the A-B bundle $\xi_{A-B}:U(1)\to R^{2*}\times U(1)\buildrel \ {\pi} \over \longrightarrow R^{2*}$ is trivial, then its gauge group ${\cal G}$ (see {\it e.g.} Refs. 1 or 3) is given by ${\cal G}=C^{\infty}(M,G)$ where $G$ is the fiber and $M$ is the base space; in our case, $G=U(1)$ and $M=R^{2*}$, then $${\cal G}=C^{\infty}(R^{2*},U(1)).\eqno{(1)}$$

Since differentiable functions are continuous, then ${\cal G}\subset C^0(R^{2*},U(1))$, and therefore the elements of ${\cal G}$ fall into different homotopy classes: $$[R^{2*},U(1)]=\{homotopy \ classes \ of \ maps \ R^{2*}\to U(1)\} \cong [S^1,S^1] \cong \pi_1(S^1)\cong
Z.\eqno{(2)}$$ So, if $f \in {\cal G}$, then there exists a unique $n \in Z$ such that $f$ is homotopic to $f_n$ ($f \sim f_n$), where $f_n(re^{i\phi})=e^{in\phi}$, and $\phi \in [0,2\pi)$, in other words, $f$ is an element of the homotopy class of $f_n$ ($f\in [f_n]$). This means that there exists a differentiable map $h:R^{2*}\times [0,1]\to U(1)$, such that $h((re^{i\phi}),0)=f(re^{i\phi})$ and $h((re^{i\phi}),1)=e^{in\phi}$. In fact, in Ref. 1 it is shown that the group of smooth homotopy classes of smooth maps from $R^{2*}$ to $U(1)$ is isomorphic to $Z$.

\

{\bf 3. Flat connections}

\

By Ref. 1, the space of flat connections on $\xi_{A-B}$ is given by the set 
$${\cal C}_0=\{A\in \Omega^1(R^{2*};u(1)), \ dA=0\}\eqno{(3)}$$
where $u(1)=iR$ is the Lie algebra of $U(1)$ and $d$ is the exterior derivative operator on $R^{2*}$.  And the action of ${\cal G}$ on ${\cal C}_0$ is given by $A \cdot f=A+f^{-1} df$, where$f^{-1}(x,y)=f(x,y)^{-1} \in U(1).$ 

\

We shall prove the following result.

\

{\it Theorem.} 

\

There is a bijection between ${\cal M}_0={\cal C}_0/{\cal G}=\{$gauge equivalence classes of flat connections on $\xi_{A-B}\}$ and $S^1$, with $length\,(S^1)=\vert e \vert$.

\

{\it Proof.} The 1-form in $R^{2*}$ which induces the A-B effect is given by $$a_0={{\Phi_0}\over {2\pi}} {{xdy-ydx}\over {x^2+y^2}}\eqno{(4)}$$ where $\Phi_0$ is the magnetic flux associated with the charge $\vert e \vert$ : ${{\Phi_0}\over {2\pi}}={{\hbar c}\over {\vert e \vert}}$, and is such that for an arbitrary flux $\Phi$ in the solenoid, $\Phi=\lambda \Phi_0$, with $\lambda \in R$; it is useful to express $\Phi_0$ in terms of the fine structure constant $\alpha$ and in the natural system of units: ${{e^2}\over {4\pi \hbar c}}=\alpha$, so $\vert e \vert =\sqrt{4\pi \alpha}$ and then ${{\Phi_0}\over {2 \pi}}={{1}\over {\sqrt{4\pi \alpha}}}\cong \sqrt{{{137}\over {4\pi}}}$.  So, $a_0={{1}\over {\sqrt{4\pi \alpha}}} {{xdy-ydx}\over {x^2+y^2}}$ and therefore $$A_0=ia_0 ={{i}\over{\sqrt{4\pi \alpha}}}{{xdy-ydx}\over{x^2+y^2}} \in {\cal C}_0.\eqno{(5)}$$ Though closed, $A_0$ is {\it not exact} since only locally, {\it i.e.} for $\phi \in (0,2\pi)$, ${{xdy-ydx}\over {x^2+y^2}}=d\phi$. 

In particular, $A_0$ generates the De Rahm cohomology (with coefficients in $iR$) of $R^{2*}$ in dimension 1 $$H^1_{DR}(R^{2*};iR)\cong H^1_{DR}(S^1;iR)=\{\lambda [A_0]_{DR}\}_{\lambda \in R}\cong R,\eqno{(6)}$$ where $[A_0]_{DR}=\{A_0+d\beta, \ \beta \in \Omega^0(R^{2*};iR)\}.$ Though  $\beta$ does not generate the most general gauge transformation of $A_0$, it gives, however, the gauge transformation defined by the composite $\hbox{exp} \circ \beta: R^{2*} \to U(1)$, \

In general, a gauge transformed of $A_0$ is of the form $A_0^{\prime}=A_0+f^{-1}df$ with $f\in {\cal G}$.  Therefore, the gauge class of $A_0$ is $$[A_0]=\{A_0+f^{-1}df\}_{f\in {\cal G}}.\eqno{(7)}$$ 

\

In order to calculate the quotient ${\cal C}_0/{\cal G}$, consider the homomorphism $\hbox{exp}_{\#} : C^{\infty}(R^{2*},iR)\longrightarrow {\cal G}=C^{\infty}(R^{2*}, U(1))$, given by $\hbox{exp}_{\#}(\beta)=\hbox{exp} \circ \beta$.

\

It is easy to see that ${\cal C}_0/{\cal G} \cong ({\cal C}_0/\hbox{Im}(\hbox{exp}_{\#}))/{\cal G}$, where the action of Im$(\hbox{exp}_{\#})$ on ${\cal C}_0$ is the action as a subgroup of
${\cal G}$, i.e., $A \cdot \hbox{exp}_{\#}(\beta)=A+\hbox{exp}_{\#}(\beta)^{-1} d\, \hbox{exp}_{\#} (\beta)$.  Since $\hbox{exp}_{\#}(\beta)^{-1} d\, \hbox{exp}_{\#}(\beta)=(\hbox{exp} \circ \beta)^{-1} (\hbox{exp} \circ \beta)d \beta$, then $A \cdot \hbox{exp}_{\#} (\beta)=A+d \beta$.  Therefore ${\cal C}_0/\hbox{Im}(\hbox{exp}_{\#})=H^1(R^{2*};iR)=\{\lambda[A_0]_{DR}\}_{\lambda
\in R}$.  The restriction imposed by the action of the full group ${\cal G}$ on the parameter $\lambda$ is obtained as follows.

\

Let $(\lambda+\sigma)A_0 \in [\lambda A_0]$, then there exists $f\in {\cal G}$ ($f$ depends on $\sigma$) such that $(\lambda+\sigma)A_0=\lambda A_0+f^{-1}df$ and therefore $f^{-1}df=\sigma A_0$ {\it i.e.} $${{1}\over{f}}{{\partial}\over {\partial x}}f={{\partial}\over {\partial x}}lnf(x,y)=-{{i\sigma}\over{\sqrt{4\pi\alpha}}} {{y}\over{x^2+y^2}},\eqno{(8)}$$ $${{1}\over{f}}{{\partial}\over {\partial y}}f={{\partial}\over{\partial y}}lnf(x,y)={{i\sigma}\over{\sqrt{4\pi \alpha}}} {{x}\over{x^2+y^2}}.\eqno{(9)}$$ Since $\int {{dx}\over{x^2+y^2}}={{i}\over{2y}}ln({{x+iy}\over{x-iy}})+const.$, $\int{{dy}\over{x^2+y^2}}={{i}\over{2x}}ln({{y+ix}\over{y-ix}})+const.^{\prime}$, and ${{y-ix}\over{y+ix}}=-{{x+iy}\over{x-iy}}$, we obtain $$lnf(x,y)={{\sigma}\over{\sqrt{16\pi \alpha}}}ln({{x+iy}\over{x-iy}})+c_1=-{{\sigma}\over{\sqrt{16\pi\alpha}}}ln({{y+ix}\over{y-ix}})+c_2$$
$$={{\sigma}\over{\sqrt{16\pi\alpha}}}ln(-{{x+iy}\over{x-iy}})+c_2={{\sigma}\over{\sqrt{16\pi\alpha}}}(ln({{x+iy}\over {x-iy}})+i\pi)+c_2,\eqno{(10)}$$  where $c_1$ and $c_2$ are constants; if $z=x+iy$ then $\bar{z}=x-iy$ and ${{x+iy}\over{x-iy}}={{z}\over {\bar{z}}}=e^{2iarg(z)}=e^{2i\phi}$, then $$lnf(x,y)={{\sigma i\phi}\over{\sqrt{4\pi\alpha}}}+c_1={{\sigma}\over{\sqrt{16\pi\alpha}}}(2i\phi+i\pi)+c_2={{i\sigma\phi}\over{\sqrt{4\pi\alpha}}}+{{i\pi\sigma}\over{\sqrt{16\pi\alpha}}}+c_2;\eqno{(11)}$$
let $\sigma=n\sqrt{4\pi\alpha}$ ($=n\vert e\vert$) with $n\in Z$, then $lnf(x,y)=in\phi+c_1=in\phi+{{i\pi n}\over{2}}+c_2$ {\it i.e.} $f(x,y)=f_n(re^{i\phi})=K_1e^{in\phi}=K_2e^{{i\pi n}\over{2}}e^{in\phi}$. Choosing $K_2=e^{-{{i\pi n}\over{2}}}$ implies $K_1=1$, and we have the solutions
$$f_n(re^{i\phi})=e^{in\phi},\eqno{(12)}$$ with $f_n(e^{i0})=f_n(e^{i2\pi})=1$. In particular, for $n=1$, we obtain $$[\lambda A_0]=[(\lambda+\sqrt{4\pi\alpha})A_0],\eqno{(13)}$$ which, as far as the classification of equivalence classes of connections and the calculation of holonomy groups is concerned, restricts the possible values of $\lambda$ to an interval of length $\sqrt{4\pi\alpha}$ which, without loss of generality, can be chosen to be $[0,\sqrt{4\pi\alpha}]\cong [0,\sqrt{{{4\pi}\over{137}}}]$ with $\sqrt{4\pi\alpha}$ identified with 0, which corresponds to the trivial connection {\it i.e.} the electromagnetic vacuum. Then, one obtains $$\pmatrix{gauge \ equivalence \ classes \cr of \ flat \ connections \ on \ \xi_{A-B} \cr } \cong {{\{[\lambda
A_0]\}_{\lambda\in [0,\sqrt{4\pi \alpha}]}}\over {0\sim \sqrt{4\pi \alpha}}} \cong {{[0,\sqrt{4\pi\alpha}]}\over {0\sim \sqrt{4\pi\alpha}}}\cong S^1. \ \ \ \ \ \ \ \ {\ q.e.d.}\eqno{(14)}$$ 

\

{\bf 4. Final remark}

\

In terms of the electric charge, ${{[0,\vert e \vert]}\over{0\sim \vert e \vert}}\cong S^1$. The ``small'' value of $\alpha$ ($\alpha\cong {{1}\over {137.04}}$) reduces the pure geometrical upper limit 1 of the interval of $\lambda$ $^{1}$, since $\sqrt{4\pi\alpha}\cong .3028<1$; then $length (S^1)=\vert e \vert$ (approximately $5.5\times 10^{-9}(erg\times cm)^{1/2}$ in the c.g.s. system of units). It is interesting that this ``length'' can be related to the Kaluza-Klein $^{4}$ length $l_{KK}$ for 5 dimensional gravity with the 5th dimension compactified in a circle giving the Maxwell field, and to the Planck length $l_P=\sqrt{G_N}$ $^{5}$, through $ l_{KK}={{2\pi l_P}\over{\vert e \vert}}$. 

\

{\bf Acknowledgments}

\

This work has been partially supported by the research grant BFM2002-03681 from the Ministerio de Ciencia y Tecnolog\'\i a and from EU FEDER funds. One of us, M. S., thanks the Spanish Ministry of Education and Culture for a sabbatical grant. We thank J. A. de Azc\'arraga for his suggestions to improve the manuscript. 

\

{\bf References} 

\

1. M. A. Aguilar, M. Socolovsky, Int. Jour. Theor. Phys. 41 (2002) 839.

\

2. Y. Aharonov, D. Bohm, Phys. Rev. 115 (1959) 485; T. T. Wu., C. N. Yang, C. N., Phys. Rev. D 12 (1975) 3845. 

\

3. J. A. de Azc\'arraga, J. M. Izquierdo, Lie Groups, Lie Algebras, Cohomology and Some Applications in Physics, Cambridge University Press, Cambridge, 1998.

\

4. Th. Kaluza, Sitz. Preuss. Akad. Wiss. K1 (1921) 966; O. Klein. O., Z. Phys. 37 (1926) 895.

\

5. M. Kaku, M., Quantum Field Theory: A Modern Introduction, Oxford University Press, New York, 1993 (Chapter 19).

\end